\documentstyle[11pt,twoside,epsf]{article}
\begin{document}

 %place pour le titre
\noindent {\bf C. R. Acad. Sci. Paris, t.xx  , S\'erie xxx , P.xx-xx ,1994}\\
\noindent Syst\`emes extragalactiques/ {\it Extragalactic Systems}
 \\
 \\
\begin{center}
\noindent {\Large {\bf Interpr\'etation de la structure de galaxies \\
entourant le Superamas Local.}}
\\
\noindent {\it Titre courant: Structure autour du Superamas Local.}
\\
\vspace{0.5cm}
Georges PATUREL, H\'el\`ene DI NELLA, Jean-No\"el TERRY et Gilles THEUREAU 
\end{center}

\noindent {\bf R\'esum\'e} --
 {\bf Interpr\'etation de la structure de galaxies entourant le Superamas Local}
 \\
Au vu des simulations num\'eriques faites par E. Praton et ses
collaborateurs, nous donnons une
interpr\'etation de la structure en cocon que nous avons observ\'ee pour la
distribution des galaxies de notre voisinage. L'effet
s'explique tr\`es simplement par les vitesses particuli\`eres engendr\'ees
par la chute sur les amas voisins. Dans cette interpr\'etation, les
structures comme le 'Cocon' (voire même comme le 'Grand-Mur') pourraient
être de simples artefacts observationnels.\\
\noindent {\it{\bf Abstract} -- 
 {\bf Interpretation of the galaxy structure surrounding the Local Supercluster.}
 \\
After numerical simulations by E. Praton, we 
interprete the cocon-like structure observed for the distribution of
galaxies around us as an effect of infall
velocities onto clusters. In this view structures like the Cocoon (or
even like the Great-Wall) could be interpreted as observationnal artefacts.
}
\vspace{1.5cm}

{\it Abridged English Version} --
In a previous paper (Di Nella et Paturel, 1994) we showed that
galaxies around us seem to form a cocoon-like structure. This
view was confirmed by the histogram
of kinematical distances (distance derived from observed radial velocity)
which shows a secondary peak at the
position of the cocoon ($r=70 Mpc$). Further, the reality of this cocoon seemed not
questionnable because a part of it is made by the
"Great-Wall" (Lapparent, Geller, Huchra, 1986) which is widely accepted as a real structure,
in agreement with theoretical predictions (Zeldovich, 1970).

Our central position inside this cocoon was somewhat amazing
even if the Local Supercluster, near the center, may open the
way for a physical explanation.
Nevertheless, our central position inside this Cocoon
was strange enough to push us to look for stronger evidence. 
Thus, we tried to find missing connections between
pieces of the Cocoon. More precisely, we search for a
density enhancement between the Pavo-Indus region and the
Perseus-Pisces one. For this purpose we measured
radial velocities using the Australian FLAIR spectrograph.
No connection has been found making the reality of the cocoon
still more questionable.

Then, a colleague call our attention to a paper by E. Praton, Mellot and McKee (1997)
presenting numerical simulations where it is shown that
any observer does have the impression to be in the center
of a Cocoon whatever his position. This effect would be caused
by the fact that distances are calculated
from observed radial velocities instead of the true cosmological
ones.

The present Note aims at giving a simple explanation of this result
using a remark by Rauzy et al. (1992) about
the distorsion of the cosmological velocity field by an
attractor (cluster or super-cluster).

Indeed, if one considers an observer O, near an attractor $A$ (cluster or super-cluster),
the cosmological radial velocity $V=H.r$ for a given galaxy G
must be increased or reduced by the radial component of
the infall velocity onto the attractor. Using the observed velocity instead
of the true cosmological velocity leads us to an observed distance
$r'$ given by Rel. 2.
Only galaxies located on the surface of the sphere of diameter $OA$
(hereafter Rauzy's sphere) 
have $r'=r$. For galaxies inside the sphere $r'>r$. For galaxies
outside the sphere $r'<r$. Then, representative points of galaxies
always move toward the surface of the sphere (figure \ref{fig4}) 
giving an artificial
density enhancement near the surface of the sphere.

However, because the infall velocity diminishes when the distance to the cluster
($GA$) increases, the density enhancement will be higher near
the cluster as shown in Fig.\ref{fig4} and Fig.\ref{fig5}. Only a part
of the sphere will be visible.

If we consider now Rauzy's spheres associated with several
superclusters around the Local Supercluster
a shell-like structure is found exactly as it has been observed
(see Fig. \ref{fig6}).

The consequences of this interpretation may change our vision
of the local universe:
The Cocoon structure (and even the Great-Wall) could be interpreted as
observationnal artefacts.
The calculation of correlation function should also be
affected by this effect.

The remaining question is the following: Are the infall
velocities large enough to created structures as strong 
as those seen? A quantitative analysis is in progress to address
this question.

\noindent ---------------------------------
 %%%%%%%%%%%%%%%%%%%%%%%%%%%%%%%%%%%%%%%%%%%%%%%%%%%%%%%%%%%%%%%%%%%%%%%%%%%
 \section{Introduction}

\begin{figure}
     \epsfxsize=12cm
 {\epsfbox[40 400 540 900 ]{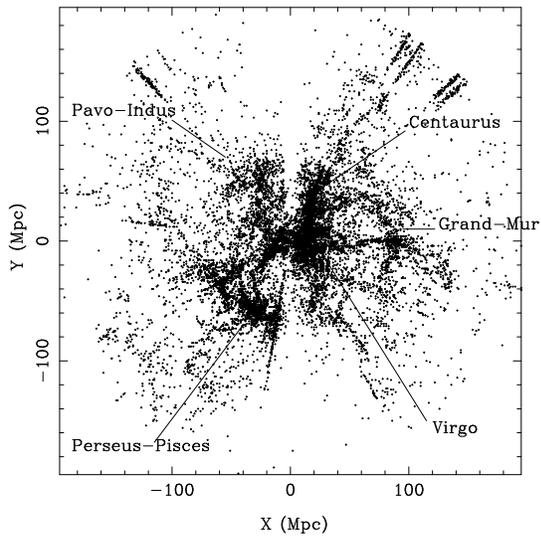}}
     \caption{
 Structure en cocon des galaxies entourant le Super-Amas-Local.
 -- {\it Cocoon-like structure of galaxies surrounding the Local Super-Cluster.}
    }
     \label{fig1}
     \end{figure}

 \noindent Dans une note pr\'ec\'edente (Di Nella et Paturel, 1994)
 nous avions montr\'e que, dans le plan contenant les principaux super-amas
 de galaxies de notre voisinage (dans un rayon de 200 Mpc environ), la
 distribution des galaxies formait une sorte d'anneau qui pouvait être la
trace d'une structure sph\'erique, le "Cocon" (figure \ref{fig1}),
 approximativement centr\'ee sur notre Galaxie. Les galaxies avaient \'et\'e
 localis\'ees dans l'espace par leurs positions en coordonn\'ees sph\'eriques
 dans le syst\`eme de coordonn\'ees "hypergalactiques" que nous avions
 d\'efini. Le rayon vecteur $r$ \'etait obtenu \`a partir de la vitesse
 radiale observ\'ee $V_{obs}$, par application de la loi de Hubble:
 $r=V_{obs}/H$, $H$ \'etant la constante de Hubble arbitrairement fix\'ee
 \`a une valeur de $75 km.s^{-1}.Mpc^{-1}$.

\begin{figure}
     \epsfxsize=7cm
     {\epsfbox[40 400 340 640 ]{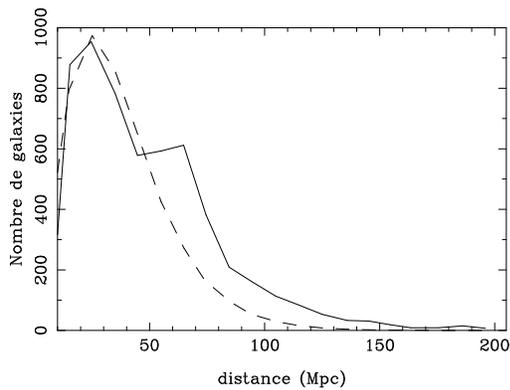}}
     \caption{
 Histogramme des distances cinématiques calculées à partir de la 
vitesse radiale observée. La courbe suit globalement 
la fonction de sélection mais un pic secondaire appara\^{\i}t pour le rayon
approximatif du cocon.
 -- {\it Histogram of kinematical distances calculated from 
observed radial velocities. The shape is the expected one
from the selection function (dashed line) but a secondary maximum appears
nearly for the radius of the cocoon.}
    }
     \label{fig2}
     \end{figure}

 \noindent L'histogramme des vitesses radiales (figure \ref{fig2}) confirmait
 pleinement l'existence de ce Cocon, car un second maximum y apparaissait
 au rayon du Cocon (environ $r=70 Mpc$). La forme g\'en\'erale de
 l'histogramme \'etait par ailleurs en excellent accord avec la forme
 pr\'evue par la fonction de s\'election.

 \noindent Un autre point confortait la description que nous donnions:
 une partie du Cocon \'etait form\'ee par le {\it Grand-Mur} d\'ecouvert
 en 1986 par Lapparent, Geller et Huchra (1986) et dont personne ne
 songeait \`a discuter la r\'ealit\'e. De plus, depuis les travaux
 de Zeldovich (1970) on s'attendait \`a trouver
 une structure alv\'eolaire dans l'Univers.

 \noindent Notre position centrale dans cette structure avait n\'eanmoins
 de quoi surprendre, m\^eme si la pr\'esence du Super-Amas-Local (super-amas
 d\'ecouvert par G. de Vaucouleurs en 1953) quasiment au centre pouvait
 justifier une interpr\'etation physique (Paturel et Di Nella, 1997) en
 invoquant une constante cosmologique non nulle.

 \noindent Pour rechercher une preuve plus concr\`ete, nous avons essay\'e de
 mettre en \'evidence une connection entre les diff\'erents arcs formant
 le Cocon. En effet, il appara\^{\i}t clairement qu'entre certaines r\'egions
 comme celle situ\'ee entre Pavo-Indus
 et Perseus-Pisces il y a une interruption de la surface
 du Cocon qui peut s'expliquer par un manque de donn\'ees. Nous
 avons donc entrepris des observations avec le spectrographe Australien
 FLAIR qui nous a permis en quelques missions de rassembler assez
 de mesures de vitesses radiales pour conclure qu'il n'y avait
 pas de surdensit\'e entre ces deux r\'egions (Di Nella et al., 1996).
 Ce r\'esultat n\'egatif rendait encore plus invraisemblable la
réalité physique du Cocon.

  \noindent Un coll\`egue allemand, H. Andernach, attira alors notre attention
  sur les travaux de Elizabeth Praton et de ses collaborateurs
(E. Praton, Mellot and McKee, 1997), qui par des simulations
  num\'eriques montraient que quelle que soit sa localisation un observateur
  avait toujours l'impression d'\^etre au centre d'une structure en
  Cocon. L'effet serait d\^u \`a ce que la distance d'une galaxie est
déduite, à travers la loi de Hubble, de la vitesse radiale {\it observ\'ee}, ce qui revient
à identifier cette vitesse radiale observée à la vitesse cosmologique
en négligeant la composante radiale de la vitesse propre de chaque galaxie.

  \noindent Le but de la pr\'esente note est de montrer que cet effet
  peut \^etre interpr\'et\'e en utilisant une remarque
 de S. Rauzy sur la perturbation du champ des vitesses par un attracteur
(amas ou super-amas).
  %%%%%%%%%%%%%%%%%%%%%%%%%%%%%%%%%%%%%%%%%%%%%%%%%%%%%%%%%%%%%%%%%%%%%%%%%%%
 \section{Perturbation de la vitesse cosmologique par les amas.}
 \subsection{Sph\`ere de Rauzy}
\begin{figure}
     \epsfxsize=6cm
     {\epsfbox[165 251 443 534]{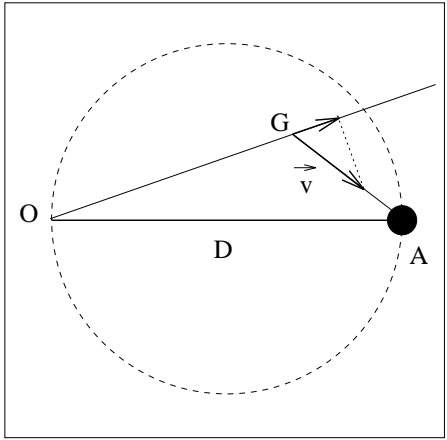}}
     \caption{Vitesse de chute d'une galaxie G vers un attracteur $A$ (amas)
 -- {\it Infall velocity of a galaxy G towards an attractor $A$ (cluster).
}
    }
     \label{fig3}
     \end{figure}

 \noindent Consid\'erons un observateur O, un attracteur $A$ (amas de galaxies) et une galaxie
 quelconque G situ\'ee \`a une distance $r$ de l'observateur
 (figure \ref{fig3}). La vitesse radiale cosmologique de cette galaxie
 sera not\'ee $V$ et sa vitesse propre de chute sur l'attracteur A sera not\'ee
 $v$. La loi de Hubble s'\'ecrit:
 \begin{equation}
        V= H.r
 \end{equation}
 o\`u $H$ est la constante de Hubble. Il est facile d'\'etablir que la distance
 radiale $r'$ calcul\'ee en utilisant la vitesse radiale observ\'ee
 au lieu de la vitesse radiale cosmologique s'\'ecrit:
 \begin{equation}
        r' = r + \frac {\vec{v}.\vec{r}}{V}
 \end{equation}
 Le lieu g\'eom\'etrique des points pour lesquels $r'=r$
 (i.e., $\vec{v}\vec{r}=0$) est la sph\`ere
 de diam\`etre $D=OA$ que nous appellerons la sph\`ere de Rauzy
 (Rauzy, Lachièze-Rey et Henriksen, 1993).

 En utilisant $r'$ \`a la place de $r$ comme on le fait n\'ecessairement
 on distord la distribution des points repr\'esentatifs des galaxies.
 Plus pr\'ecis\'ement:
\begin{itemize}
\item Si $\vec{v}\vec{r}>0$  (points int\'erieurs \`a la sph\`ere de Rauzy)
alors $r'>r$. Les points seront déplac\'es en direction de la surface de la sph\`ere.
Un point $G_1$ sera plac\'e en $G'_1$ (voir figure \ref{fig4})

\item Si $\vec{v}\vec{r}<0$  (points ext\'erieurs \`a la sph\`ere de Rauzy)
alors $r'<r$. Les points seront aussi déplac\'es en direction de la surface
de la sph\`ere. Un point $G_2$ sera plac\'e en $G'_2$.
\end{itemize}

\noindent En conclusion, la sph\`ere de Rauzy devrait
appara\^{\i}tre comme une surdensit\'e.

 \subsection{Variation de la surdensit\'e sur la sph\`ere de Rauzy}
\begin{figure}
     \epsfxsize=6cm
     \hbox{\epsfbox[165 251 443 534]{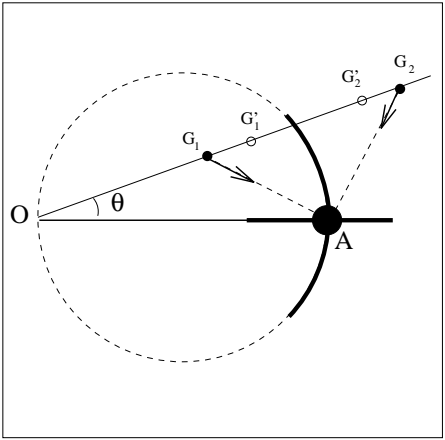}}
     \caption{Une galaxie en G est placée à la position G'. Une surdensité
appara\^{\i}t près de l'amas de galaxies.
 -- {\it A galaxy in G is plotted in G'. The density enhancement appears
near the cluster.
}
    }
     \label{fig4}
     \end{figure}

 \noindent La surdensit\'e apparente au
 voisinage de la sph\`ere de Rauzy ne sera pas la m\^eme partout.
 La vitesse de chute sur un amas est
 de la forme $v \propto d^{1-\gamma}$, o\`u $\gamma$ est une constante
 et $d=GA$ la distance \`a l'attracteur. Pour un angle de vis\'ee donn\'e $\theta$ 
 il y a une distance minimum limite ($d_{lim}=D.\sin \theta$).
 Donc n\'ecessairement la surdensit\'e
 au voisinage de la sph\`ere sera forte pr\`es de l'amas et faible
près de l'observateur
  \footnote{Dans notre cas, les galaxies proches de l'observateur seront
  \'egalement perturb\'ees par le Super-Amas-Local}
 comme l'illustre la figure \ref{fig4}, d'autant plus que 
la densité réelle est naturellement plus élevée au voisinage de l'amas
que dans le champ générale. Finalement,
seule la partie de la sphère de Rauzy proche de l'amas sera visible.

\begin{figure}
     \epsfxsize=6cm
 {\epsfbox[60 425 350 720 ]{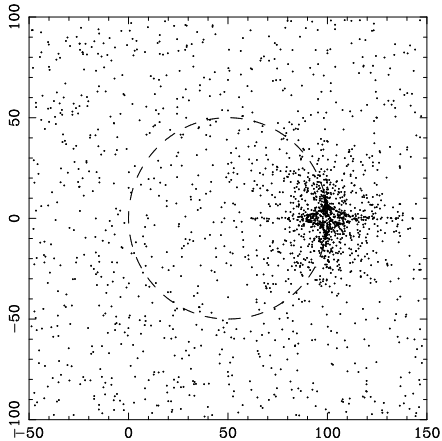}}
     \caption{Une simulation faite avec des valeurs réalistes montre l'effet 
de surdensité en arc de cercle au voisinage de l'amas ainsi que l'effet 
d'étirement dans la direction de l'amas ('Doigt de Dieu'). L'observateur
est supposé être au point de coordonnées ($0,0$) et l'amas au point ($100,0$).
--{\it A simulation with realistic values shows the effect of curved density enhancement
near the cluster and the 'Finger of God' effect. The observer is located
at ($0,0$) and the cluster at ($100,0$).
 }
    }
     \label{fig5}
     \end{figure}

Pour illustrer cet effet nous avons réalisé
une simulation numérique. L'observateur étant à la position ($x=0$, $y=0$)
nous avons pris une distribution uniforme de
1000 galaxies (galaxies de champ) superposée à une distribution gaussienne de 1000 galaxies 
(amas) centrée à la position supposée de l'amas ($x=100$, $y=0$) et d'écart-type 
 $\sigma=20$.
Nous avons adopté une vitesse de chute de la forme (Peebles, 1976):
 \begin{equation}
        v= \frac {C}{d^{\gamma -1}}
 \end{equation}
avec C=1300 et $\gamma=1.5$ nous avons obtenu la figure \ref{fig5}
qui montre bien l'effet. Les valeurs numériques adoptées
donneraient pour la chute de notre Groupe Local sur l'amas Virgo ($d \approx 18Mpc$)
une vitesse de $300 km.s^{-1}$, ce qui est conforme aux observations.

 \noindent Notons que lorsque la distance $d$ tend vers z\'ero, la vitesse de chute
 augmente beaucoup et la surdensit\'e peut appara\^{\i}tre sur un grand domaine
 de distance autour de l'amas. Cet effet conduit aux alignements
 appel\'es "Doigt de Dieu" (parce qu'ils pointent vers l'observateur) mais dont
 l'origine avait d\'ej\`a \'et\'e comprise.

 \subsection{Effet de plusieurs super-amas}
\begin{figure}
     \epsfxsize=6cm
     \hbox{\epsfbox[146 239 462 553]{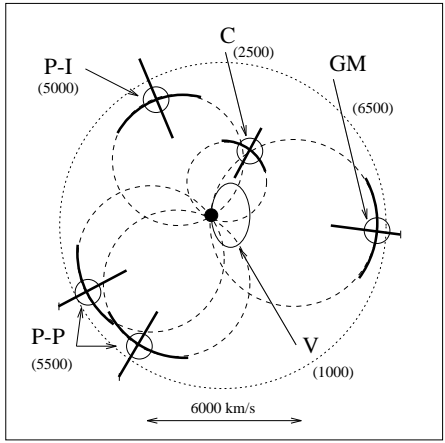}}
     \caption{
Plusieurs surdensités en arc de cercle donnent l'impression à
l'observateur d'être au centre d'une structure sphérique parce que les amas
les plus riches apparaissent tous approximativement à la même distance de nous
à cause de la fonction de sélection qui limite la profondeur d'exploration
(voir figure2). Pour chaque concentration repérée sur la figure 1 nous
donnons la vitesse radiale approximative en $km.s^{-1}$.
 -- {\it Several curved density enhancements look like a spherical
structure centered on the observer because the richest clusters all appear
nearly at the same distance from us due the selection function which
limits the depth of exploration (see figure 2). For each galaxy cluster
quoted in figure 1 we give the approximate radial velocity in $km.s^{-1}$.}
}
     \label{fig6}
     \end{figure}

 \noindent Nous avions défini le plan hypergalactique comme le plan le plus peupl\'e
 en galaxies. Au moins cinq amas ou super-amas s'y trouvent localis\'es: le
 Super-Amas-Local, le super-amas de Pavo-Indus, le super-amas de
 Perseus-Pisces, le super-amas de Coma et l'amas de Centaurus. Plusieurs d'entre eux
 (PI, P-P et Coma) ont une vitesse comprise entre $5\ 000$
 et $7\ 000 km.s^{-1}$. Les surdensit\'es sur les sph\`eres de Rauzy
 associ\'ees \`a ces super-amas donnent une interpr\'etation plausible
 du Cocon comme la figure \ref{fig6} l'illustre.

 %%%%%%%%%%%%%%%%%%%%%%%%%%%%%%%%%%%%%%%%%%%%%%%%%%%%%%%%%%%%%%%%%%%%%%%%%%%
 \section{Conclusions}
 \noindent Plusieurs cons\'equences intéressantes
d\'ecoulent de l'interpr\'etation ci-dessus:
 \begin{itemize}
 \item La structure en Cocon serait un artefact observationnel. Notre position centrale
serait un effet apparent qui s'expliquerait de manière très naturelle sans qu'il soit 
nécessaire d'invoquer des hypothèses nouvelles.
 \item La surdensité du "Grand-Mur" qui est une structure plus simple 
formant une partie du Cocon serait 
renforcée par l'effet décrit ci-dessus et pourrait, à la limite, être vue aussi
comme un artefact observationnel.
 \item Les fonctions de corr\'elation calcul\'ees habituellement \`a partir
 des distances d\'eduites de la vitesse radiale observ\'ee devraient
 \^etre affect\'ees. 
 \end{itemize}

 \noindent La derni\`ere question qui se pose est la suivante:
 Les vitesses particuli\`eres engendr\'ees par l'attraction gravitationnelle
 des amas ou super-amas sont-elles suffisamment grandes pour expliquer
 des structures aussi marqu\'ees que le Cocon ou le Grand-Mur.
 Une analyse quantitative est en cours mais les simulations de
 Elizabeth Praton semblent promettre une r\'eponse affirmative.
 %%%%%%%%%%%%%%%%%%%%%%%%%%%%%%%%%%%%%%%%%%%%%%%%%%%%

\vspace{1.5cm}

\small
 REFERENCES BIBLIOGRAPHIQUES \\

DE LAPPARENT V., GELLER M., HUCHRA J., 1986, {\it A slice of the Universe}, Astrophys. J. 302, p.L1-L5

DE VAUCOULEURS G., 1953, {\it Evidence for a local Supergalaxy}, Astronomical Journal, 58, p.30-32

DI NELLA H., PATUREL G., 1994 {\it Structure à grande échelle de l'Univers jusqu'à la distance de 200 Mpc}, C. R. Acad. Sci. Paris, t.319, Série II, p.57-62

DI NELLA H., COUCH W., PATUREL G.,  PARKER Q., 1996, {\it Are the Perseus-Pisces chain and the Pavo-Indus wall connected?}, Mon. Not. R. Astron. Soc. 283, p.367-380

PATUREL G., DI NELLA H., 1995, {\it Galaxy Distribution Around The Local Supercluster}, Astro. Lett. \& Communications 31, p.337-340

PEEBLES P.J.E., 1976, {\it The peculiar velocity field in the Local Supercluster}, Astrophys. J. 205, p.318-328 

PRATON E.A., MELOTT A.L., MCKEE M.Q., 1997, {\it The Bull's-Eye Effect: Are Galaxy Walls Observationally Enhanced ?}, Astrophys. J. 479, p.L15-18

RAUZY S., LACHIEZE-REY M., HENRIKSEN R.N., 1992, {\it Detecting non-Hubble velocity fields in the universe}, Astron. Astrophys. 256, p.1-9

ZEL'DOVICH Y.B., 1970, {\it Gravitational Instability: An Approximate Theory for Large Density Perturbations}, Astron. Astrophys. 5, p.84-89

\vspace{1cm}

{\it
\hspace {2cm} \noindent PATUREL G.: Observatoire de Lyon, 69561 Saint-Genis Laval CEDEX, France

\hspace {4cm}  \noindent TEL: 04.78.86.83.83  FAX: 04.78.86.83.86  E-MAIL: patu@obs.univ-lyon1.fr

\hspace {4cm}  \noindent (PATUREL G. corrigera les épreuves)

\hspace {2cm}  \noindent DI NELLA H.: Observatoire de Lyon, 69561 Saint-Genis Laval CEDEX, France

\hspace {4cm}  \noindent TEL: 04.78.86.83.83  FAX: 04.78.86.83.86  E-MAIL: courtois@obs.univ-lyon1.fr

\hspace {2cm}  \noindent TERRY J.N.: Observatoire de Lyon, 69561 Saint-Genis Laval CEDEX, France

\hspace {4cm}  \noindent TEL: 04.78.86.83.83  FAX: 04.78.86.83.86  E-MAIL: jnterry@obs.univ-lyon1.fr

\hspace {2cm}  \noindent THEUREAU G.: Observatoire de Paris, 92195 Meudon CEDEX, France

\hspace {4cm}  \noindent TEL: 01.45.07.76.04  FAX: 01.45.07.79.39  E-MAIL: theureau@obspm.fr

}

 \end{document}